\documentclass[aps,prl,reprint,nofootinbib]{revtex4-1}
\def\prd{{Phys.~Rev.~D}}        % Physical Review D
\usepackage{amsmath,amstext}
\usepackage[T1]{fontenc}
\usepackage{amssymb}
\input{epsf}
\usepackage{graphicx}
\usepackage{ae,aecompl}

\usepackage{hyperref}
\usepackage{amsmath}
\usepackage{amssymb}
\usepackage{mathtools}
\usepackage{bm}
\usepackage{cleveref}
\usepackage{tensor}
\usepackage{braket}
\usepackage{enumitem}
\usepackage{mhchem}

\usepackage{color}

\newcommand{\spc}{\quad \quad \quad}

\def\be{\begin{equation}}
\def\ee{\end{equation}}
\def\beq{\begin{eqnarray}}
\def\eeq{\end{eqnarray}}

\makeatletter 
    
\renewcommand\onecolumngrid{% <<<<<<
\do@columngrid{one}{\@ne}%
\def\set@footnotewidth{\onecolumngrid}% <<<<<<<<<<<<<<<<
\def\footnoterule{\kern-6pt\hrule width 1.5in\kern6pt}%
}

\renewcommand\twocolumngrid{% <<<<<<
        \def\footnoterule{% restore rule
        \dimen@\skip\footins\divide\dimen@\thr@@
        \kern-\dimen@\hrule width.5in\kern\dimen@}
        \do@columngrid{mlt}{\tw@}
}%

\makeatother    
\begin{document}
\title{Thermodynamic stability implies causality}
\author{L.~Gavassino, M.~Antonelli \& B.~Haskell}
\affiliation{Nicolaus Copernicus Astronomical Center, Polish Academy of Sciences, ul. Bartycka 18, 00-716 Warsaw, Poland}

\begin{abstract}
The stability conditions of a relativistic hydrodynamic theory can be derived directly from the requirement that the entropy should be maximised in equilibrium. Here we use a simple geometrical argument to prove that, if the hydrodynamic theory is stable according to this entropic criterion, then localised perturbations to the equilibrium state cannot propagate outside their future light-cone. In other words, within relativistic hydrodynamics, acausal theories must be thermodynamically unstable, at least close to equilibrium. We show that the physical origin of this deep connection between stability and causality lies in the relationship between entropy and information. Our result may be interpreted as an ``equilibrium conservation theorem'', which generalizes the Hawking-Ellis vacuum conservation theorem to finite temperature and chemical potential.  
\end{abstract} 

\maketitle

\textit{Introduction -} A hydrodynamic theory is said to be stable if small deviations from the state of global thermodynamic equilibrium do not have the tendency to grow indefinitely, but remain bounded over time. It is said to be causal if signals do not propagate faster than light.  Every hydrodynamic theory should guarantee the validity of these two principles, the former arising from the definition of equilibrium as the state towards which dissipative systems evolve as $t \rightarrow +\infty$, the latter arising from the principle of relativity (if signals were superluminal, there would be a reference frame in which the effect precedes the cause). Whenever a new theory is proposed, it needs to pass these two tests, to be considered reliable. To date, these properties have been mostly studied as two distinct, disconnected features of the equations of the theory, to be discussed separately. Intuitively, this approach seems natural, as stability and causality are two principles which pertain to two complementary branches of physics: thermodynamics \cite{PrigoginebookModernThermodynamics2014} and field theory \cite{Peskin_book}. 

However, in reality these two features appear to be strongly correlated. Divergence-type theories are causal if and only if they are stable \cite{GerochLindblom1990} while Israel-Stewart theories are causal if they are stable \cite{Hishcock1983,OlsonLifsh1990}. \citet{Geroch_Lindblom_1991_causal} analysed a wide class of causal theories for dissipation and found that many causality conditions have an important stabilising effect. Finally, \citet{BemficaDNDefinitivo2020} recently proved a theorem, according to which, if a strongly hyperbolic theory is stable in the fluid rest-frame, and it is causal, then it is stable in every reference frame, formalising a widespread intuition \cite{DenicalStability2008,PuKoide2010}. All these results suggest the existence of an underlying physical mechanism connecting causality and stability. Discovering it would lead to a complete change of paradigm. In fact, it would provide a new insight into the physical meaning of the (usually complicated) mathematical structure which ensures causality. Furthermore, it would importantly simplify the (usually tedious) job of testing both causality and stability, maybe reducing one to the other. 

To date, a ``fully explanatory'' mechanism connecting causality and stability has never been proposed. In fact, such a connection is usually found a posteriori, by direct comparison between the two distinct sets of conditions (as in \cite{Hishcock1983}), or through complicated mathematical proofs, as in \cite{BemficaDNDefinitivo2020}. The goal of this letter is to finally explain simply the relationship between causality and stability. We prove, with a geometrical argument, that if a theory is \textit{thermodynamically} stable, namely if the entropy is maximised at equilibrium (see \citet{GavassinoGibbs2021}), it is also causal, close to equilibrium\footnote{We restrict our attention to linear causality, namely to the requirement that the retarded Green's function of the linearised problem should vanish outside the future light-cone \cite{Susskind1969}.}. We show that the key to understand this result from a physical perspective is the underlying relationship between entropy and information. Furthermore, we explain why causality alone does \textit{not} imply stability (see e.g. \cite{Susskind1969,MasoudHall2021}), but one needs at least to prove stability in a particular reference frame (in agreement with \cite{BemficaDNDefinitivo2020}). 

We adopt the signature $(-,+,+,+)$ and we work in natural units $c=k_B=1$.

\textit{Thermodynamic stability -} Under which conditions is a relativistic fluid thermodynamically stable? Consider a fluid ``F'' that is in contact with a heat-particle bath ``H''. Assume that the total system ``$\text{fluid}+\text{bath}$'' is isolated and evolves spontaneously from a state $1$ to a state $2$. Then the total entropy should not decrease (given a quantity $A$, we call $\Delta A:= A_2-A_1$):
\begin{equation}
\Delta S_{\text{tot}}=\Delta S_\text{F} + \Delta S_\text{H} = \Delta S_F + \int_1^2 dS_\text{H}  \geq 0 \, .
\end{equation}
If $Q^I$ are the relevant conserved charges of the system, e.g. baryon number and four-momentum \cite{Peskin_book}, we can write $dS_\text{H} =-\alpha_I^\text{H} dQ^I_\text{H}$, where $\alpha_I$ are the thermodynamic conjugates of $Q^I$ and we are adopting Einstein's convention for the index $I$. Considering that $dQ^I_\text{H} =-dQ^I_\text{F}$ (charge conservation), and that the bath is \textit{defined} as a system that is so large that $\alpha_I^\text{H}=\text{const}=:\alpha_I^\star$ in any interaction with F \cite{Pathria2011,Termo,GavassinoTermometri}, we find that ($\alpha_I^\star$ are constants)
\begin{equation}\label{dittuto}
\Delta S_{\text{tot}}= \Delta (S_\text{F} +\alpha_I^\star Q^I_\text{F}) \geq 0.
\end{equation}
This implies that the equilibrium state of F is the state that maximises the functional $\Phi=S_\text{F} +\alpha_I^\star Q^I_\text{F}$ for unconstrained variations \cite{Callen_book,landau5,Pathria2011,Stuekelberg1962,Termo,Israel_2009_inbook,
GavassinoTermometri,Ottinger1997}. Hence, for an arbitrary space-like 3D-surface $\Sigma$ which extends over the support of F, we need to require that
\begin{equation}\label{Esigma}
\begin{split}
& E[\Sigma]:= -\delta \Phi[\Sigma] = \int_{\Sigma}  E^a  n_a \, d \Sigma \geq 0 \\
& \text{with} \quad E^a =-\delta (s^a +\alpha^\star_I J^{Ia})=-\delta s^a -\alpha^\star_I \, \delta J^{Ia} \, , \\
\end{split}
\end{equation}
where $s^a$ is the fluid's entropy current, $J^{Ia}$ are the currents whose fluxes are $Q^I_\text{F}$, and ``$\, \delta \,$'' is an arbitrary \textit{finite} perturbation from the equilibrium state. In most applications, $E^a$ may be truncated to second order in the perturbations $\delta \varphi_i$ to the hydrodynamic fields (like the fluid four-velocity and the temperature field).

Let us list the most important properties of $E^a$:
%Consider a relativistic fluid located in a stationary (fixed) background spacetime \footnote{The stationarity of the spacetime guarantees the conservation of the total energy of the fluid associated with the Killing field.}. Given a selection of the macroscopic fields ($\varphi_i$) which carry information about the local state of the fluid (e.g. the fluid four-velocity and the temperature field) we consider two solutions of the hydrodynamic equations, $\varphi_i$ and $\varphi_i + \delta \varphi_i$. The first is the state of global thermodynamic equilibrium, while $\delta \varphi_i$ is a small perturbation which conserves the total energy and any other conserved quantity (such as the total baryon number). 
%
%Assume that the fluid is thermodynamically stable. Then, following \cite{GavassinoGibbs2021}, we know that, for an arbitrary \textit{space-like} Cauchy 3D-surface $\Sigma$, the quantity
%\begin{equation}\label{differisco}
%E[\Sigma] := S[\Sigma,\varphi_i] - S[\Sigma,\varphi_i +\delta \varphi_i],
%\end{equation}
%quantifying the entropy difference between the equilibrium and the perturbed state, can be written as the flux
%\begin{equation}\label{Esigma}
%E[\Sigma] = \int_{\Sigma}  E^a  n_a \, d \Sigma 
%\end{equation}
%of a four-current $E^a=E^a[\varphi_i, \, \delta \varphi_i]$, which has the following properties: 
\begin{enumerate}
\item[(i) -] For any unit vector $n^a$, time-like and past-directed ($n^a n_a=-1$, $n^0<0$), we have
\begin{equation}\label{eaea}
E^a n_a \geq 0.
\end{equation}
\item[(ii) -] For the same $n^a$ as in (i), $E^a n_a=0$ on any point where the perturbation to every observable is zero, and only on these points. 
\item[(iii) -] The four-divergence of $E^a$ is non-positive:
\begin{equation}
\nabla_a E^a  \leq 0.
\end{equation}
\end{enumerate}
The first property follows from $E[\Sigma]\geq 0$, which must hold for any space-like 3D-surface $\Sigma$ covering F \footnote{Note that, in equation \eqref{Esigma}, $\Sigma$ does not necessarily cover all the space. For example, if we define F and H to be two portions of a same fluid (with H infinitely larger than F), $\Sigma$ must cover \textit{only} the support of F, and not that of H. Since the distinction between F and H is ultimately a convention, we need to require $E \geq 0$ for \textit{any} $\Sigma$ space-like, leading to condition (i). See also \cite{landau5} \S 21 ``\textit{Thermodynamic inequalities}'' for a similar argument.}. Note that, the vector $n^a=n^a[\Sigma]$ appearing in \eqref{Esigma} is the unit normal to $\Sigma$, which is time-like past-directed \cite{MTW_book}. The second property follows from the definition of $E^a$, and from the assumption that the equilibrium state is unique. The third property follows from \eqref{dittuto}. Conditions (i,ii,iii) imply that $E$ is a non-increasing ``square-integral norm'' of the perturbation $\delta \varphi_i$, enforcing the Lyapunov-stability of the equilibrium state \cite{lasalle1961stability,Prigogine1978,GavassinoLyapunov_2020}. In the Supplementary Material we show that (i,ii,iii) are mathematically equivalent to the Gibbs stability criterion \cite{GavassinoGibbs2021}. 

The criterion for \textit{thermodynamic stability} described above is a sufficient condition for \textit{hydrodynamic stability}, but contains more information than a hydrodynamic stability analysis: while the latter is a dynamical property of the field equations (an on-shell criterion\footnote{By \textit{on-shell}  we mean ``along solutions of the field equations'', by \textit{off-shell} we mean ``independently from the field equations''.}), the former is a property of the constitutive relations (it must be respected also off-shell). In fact, thermodynamic stability also implies stability to thermodynamic fluctuations, whose probability distribution \cite{landau5,Pathria2011},
\begin{equation}
\mathcal{P}[\delta \varphi_i] \propto e^{\delta \Phi[\Sigma,\delta \varphi_i]} = e^{-E[\Sigma,\delta \varphi_i]}, 
\end{equation}
must be peaked at $\delta \varphi_i=0$, leading to conditions (i,ii).

To see the difference between hydrodynamic and thermodynamic stability, consider the case of a perfect fluid, whose current $E^a$ is \cite{Hishcock1983,GavassinoGibbs2021,Stuekelberg1962}
\begin{equation}\label{Eaperf}
\begin{split}
TE^a = \, & \dfrac{u^a}{2} (\rho +p) \delta u^b \delta u_b +  \delta u^a \delta p  \, +\\ 
& \dfrac{u^a}{2} \bigg[\dfrac{1}{c_s^{2}} \, \dfrac{(\delta p)^2}{\rho +p} + \dfrac{nT}{c_p} \, (\delta \sigma)^2 \bigg]+\mathcal{O}[(\delta \varphi_i)^3], \\
\end{split} 
\end{equation}
where $u^a$, $n$, $T$, $\rho$, $p$, $\sigma$, $c_s$ and $c_p$ are fluid velocity, particle density, temperature, energy density, pressure, entropy per particle, speed of sound and specific heat at constant pressure (quantities without ``$\, \delta \,$'' are evaluated at equilibrium). Conditions (i,ii) produce the thermodynamic inequalities (assuming $n,T >0$)
\begin{equation}
0 <  c_s^2 \leq 1 \quad \quad \rho +p > 0  \quad \quad c_p > 0.
\end{equation} 
A positive $c_p$ guarantees stability to heat transfer. However, since a perfect fluid does not conduce heat, the inequality $c_p >0$ is invisible to a hydrodynamic stability analysis. On the other hand, thermodynamic stability implies stability also to virtual processes \cite{Callen_book}, which become real when thermal fluctuations are included in the description \cite{KovtunStickiness2011,TorrieriCrooks2020}, or when we couple the fluid with other fluids \cite{GavassinoRadiazione} or heat baths \cite{Termo,GavassinoTermometri}.

Finally, it is also relevant to mention that, in ideal-gas kinetic theory, $E^a$ \textit{always} obeys conditions (i,ii,iii), and is given by \cite{GavassinoGibbs2021,Israel_Stewart_1979} ($\varepsilon$ is $+1$ for bosons and $-1$ for fermions)
\begin{equation}\label{kinetco}
E^a = \int \dfrac{(\delta f)^2 p^a}{2f (1 +\varepsilon f)} \dfrac{d^3 p}{p^0} + \mathcal{O}[(\delta f)^3],
\end{equation}
where $f=f(x,p)$ is the invariant distribution function, counting the number of particles in a small phase-space volume centered on $(x,p)$ \footnote{Equation \eqref{kinetco} is written local inertial coordinates, with units such that $h^3=g_s$, where $h$ is Planck's constant and $g_s$ the spin degeneracy.}. Hence, for ideal gases, the conditions of thermodynamic stability (i,ii,iii) are also a criterion of consistency with the kinetic description.

\begin{figure}
\begin{center}
\includegraphics[width=0.5\textwidth]{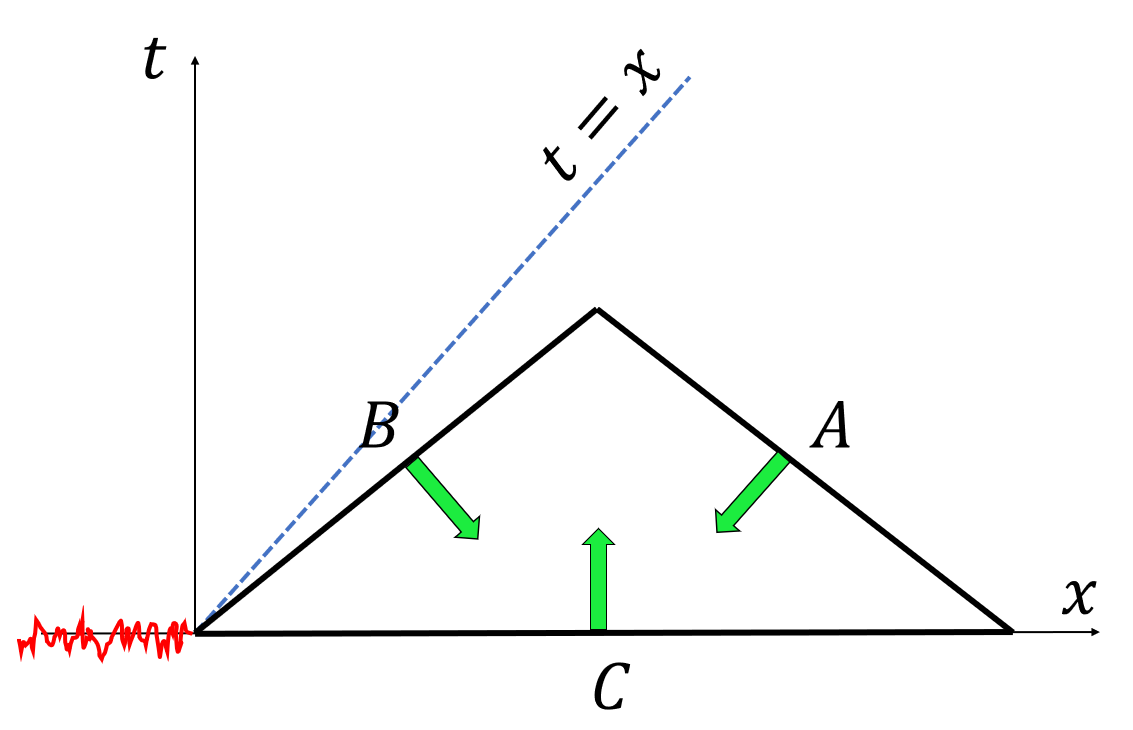}
	\caption{Visualization of the geometric argument. The initial perturbation (in red) is located to the left of the origin. Causality requires it to stay confined in the $t \geq x$ half-plane. We build the triangle $ABC$ in a way that all its edges are space-like. $B$ and $C$ intersect at the origin. We can regulate $B$ to be arbitrarily close to the line $t=x$. The green arrows are a Euclidean representation of the unit normal vectors to the edges and are taken inward-pointing, consistently with our choice of metric signature \cite{Hawking1973}.}
	\label{fig:fig}
	\end{center}
\end{figure}

\textit{The argument for causality -} Our goal is to show that the conditions (i,ii,iii) imply causality. We work, for clarity, in 1+1 dimensions, on a scale that is assumed sufficiently small that we can neglect the gravitational field.  The generalization to 3+1 dimensions (and curved space-time) is presented in the Supplementary Material. Working in an inertial coordinate system $(t,x)$, we consider a perturbation $\delta \varphi_i$ that is initially confined on the semi-axis $x \leq 0$, namely
\begin{equation}\label{solosotto}
\delta \varphi_i(0,x)=0  \spc \forall \, x > 0.
\end{equation}
We apply the Gauss theorem to the triangle $ABC$ shown in figure \ref{fig:fig} and use condition (iii):
\begin{equation}\label{ILGAUSS}
E[A]+E[B]+E[C] = \int_{(\text{triangle})} \hspace*{-1cm} \nabla_a E^a \, dt \, dx \leq 0.
\end{equation}
The 1D surfaces $A$, $B$ and $C$ are all space-like, so that their unit normal vector must be taken inward-pointing \cite{Hawking1973}. Combining \eqref{solosotto} with condition (ii) we obtain $E[C]=0$. Furthermore, since the unit normals to $A$ and $B$ are time-like past-directed, we can use (i) to show that $E[A]$ and $E[B]$ are non-negative, so that \eqref{ILGAUSS} implies
\begin{equation}
E[A]=E[B]=0.
\end{equation}
But this implies, recalling (ii), that $\delta \varphi_i$ must be zero on all the sides of the triangle. Since we can make the triangle arbitrarily long ($A$ and $C$ may extend to $x=+\infty$) and the side $B$ may be arbitrarily close to the line $t=x$ (without crossing it, because $B$ must be space-like), we finally obtain
\begin{equation}
\delta \varphi_i (t,x)=0  \quad \quad \text{for}  \quad \quad x>t. 
\end{equation}
This shows that no perturbation can propagate outside the light-cone, hence linear causality\footnote{Our argument tells us nothing about how signals propagate inside a region where $\delta \varphi_i$ is \textit{already} different from zero. For this reason, our argument can serve only to prove \textit{linear} causality, as defined in \cite{Susskind1969}.}.

\textit{Physical interpretation -} To be able to understand the physical meanig of the argument above, we need first to have an intuitive interpretation of $E^a$. 
%We start by recalling that Boltzmann's formula $S=\ln \Gamma$, where $\Gamma$ is the number of microscopic realizations of a given macrostate, implies that we may interpret $S$ as a measure of our ignorance about the microstate in which the system is.

Within the usual interpretation of entropy as uncertainty, in the sense that $S_{\text{tot}}$ reflects our ignorance, interpreted as lack of information \cite{JaynesInformation1957}, about the exact system's microstate (recall Boltzmann's formula $S_{\text{tot}}=\ln \Gamma$, where $\Gamma$ is the number of microscopic realizations of a given macrostate), equation \eqref{dittuto} implies
\begin{equation}
E =  \binom{\text{ Ignorance at }}{\text{ equilibrium }} - \binom{\text{ Ignorance in the }}{\text{ perturbed state }} .
\end{equation}
Hence, $E$ is the net information carried by the perturbation. The Gibbs stability criterion ($E \geq 0$), then, is the statement that any perturbation increases our knowledge about the microstate. 
Now, if we look at equation \eqref{Esigma} and invoke condition (ii), it follows that we can identify $E^a$ with the current of information 
%\footnote{To avoid confusion, we remark that, chosen a fluid element $X$, $E^a[X]$ does \textit{not} describe the flow of information relative to the microstate of $X$ (this would be given by $s^a[X, \,\varphi_i]-s^a[X, \, \varphi_i + \delta \varphi_i]$, where $s^a$ is the entropy current). Instead, $E^a[X]$ quantifies the flow of information relative to the microstate \textit{of the whole fluid} that we can extract by performing measurements on $X$.} 
transported by the perturbation (see Supplementary Material for a direct proof). In fact, if $E^a=0$ in a given region of space $\mathcal{R}$, then the average value of any observable on $\mathcal{R}$ coincides with the microcanonical average (i.e. the equilibrium value). Since the microcanonical ensemble assigns equal a priori probability to every microstate, there is no information in $\mathcal{R}$. 

Now that we have an interpretation of $E^a$, let us examine conditions (i) and (iii). The latter is the second law of thermodynamics, as seen from the point of view of information theory: our initial information about the microstate of the system can only be lost (or transported from one place to another) in time, but never created, because all the initial conditions tend, as $t \rightarrow +\infty$, to the same final macrostate (the equilibrium). However, the most interesting condition for us is (i): it is easy to show that imposing (i), namely that the density of information is non-negative in any frame, is equivalent to requiring that $E^a$ is time/light-like future-directed, namely
\begin{equation}
E^a E_a \leq 0  \spc E^0 \geq 0.
\end{equation}
This is where the contact with causality is established. In fact, if information is transported by a non-space-like four-current, it propagates along causal trajectories and cannot exit the light-cone (namely, no perturbation can transport information faster than light). This result may be seen as the finite-temperature analogue of the Hawking Ellis vacuum conservation theorem \cite{Hawking1973,CarterVacuum2003}. It establishes that \textit{information} (in their case \textit{energy}) is not spontaneously formed in an \textit{equilibrium} (in their case \textit{empty}) region and cannot enter it from outside its causal past. In this analogy, the Gibbs stability criterion plays the role of the dominant energy condition.

\textit{The inverse argument -} It is natural to ask whether we can reverse the argument and show that causality implies stability. This is in general not true (see e.g. \cite{Susskind1969,MasoudHall2021}). In fact, let us assume that we still have an information current $E^a$, defined by equation \eqref{Esigma}, and that conditions (ii) and (iii) are valid (they are typically ensured by construction when there is an entropy current). The causality requirement reduces to imposing that $E^a$ is time/light-like, but this does not specify its orientation. It might be the case that $E^a$, for some configurations, is past-directed, generating instability. Thus, in general
\begin{equation*}
(\text{Causality})  \spc \Longrightarrow \hspace*{-0.5cm} / \spc \text{(i)}.
\end{equation*}
However, to fix the orientation we only need to assume that there is a preferred reference frame in which $E^0 \geq 0$ $\forall \, \delta \varphi_i$. It is natural, and it usually simplifies the calculations, to take this reference frame to be aligned with the equilibrium inverse-temperature four-vector $\beta^a$, which always exists, is unique and is time-like future-directed \cite{BecattiniStress2012,Becattini2016Beta,GavassinoTermometri}. Hence, we can conclude that
\begin{equation*}\label{capusta}
(\text{Causality}) \quad  + \quad  (E^a \beta_a \leq 0)  \quad \quad \Longrightarrow  \quad \quad \text{(i)},
\end{equation*}
which is consistent with the more general theorem of \citet{BemficaDNDefinitivo2020}.

\begin{figure}
\begin{center}
\includegraphics[width=0.5\textwidth]{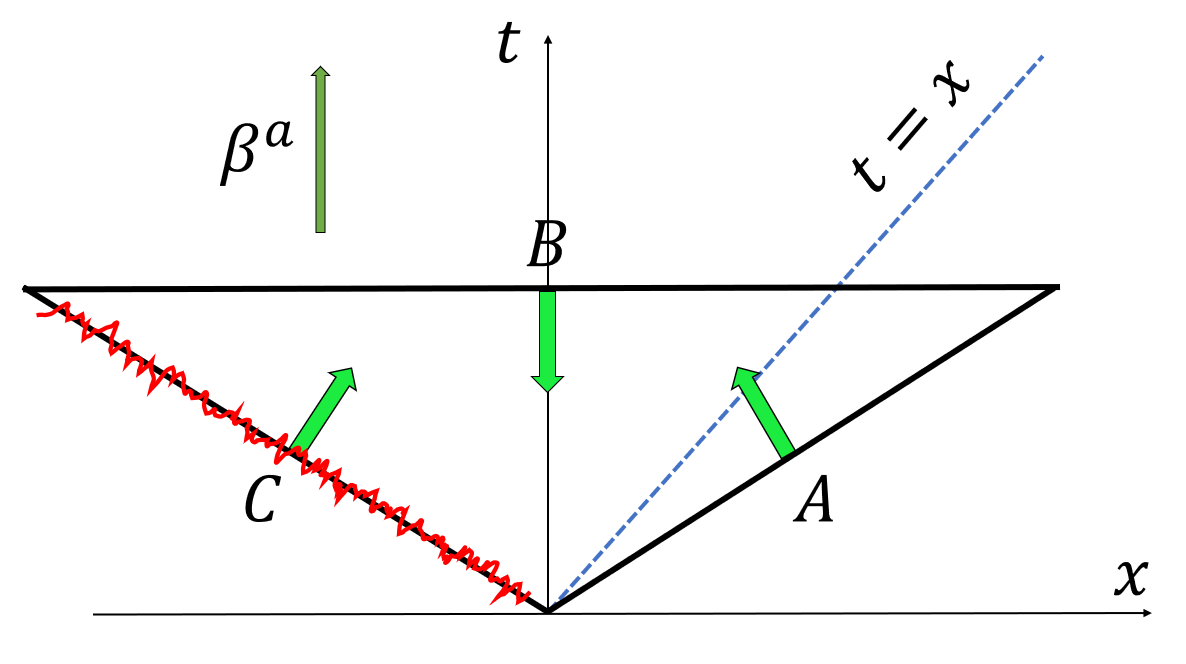}
	\caption{Visualization of the geometric argument for the theorem of \citet{BemficaDNDefinitivo2020}. All the edges of the triangle $ABC$ are space-like. We create an arbitrary initial perturbation (in red) on the side $C$. Since $A$ is outside the causal future of $C$, we are free to set the perturbation to zero on $A$. The inverse temperature four-vector $\beta^a$ (dark green) is aligned with the t-axis. The light green arrows are a Euclidean representation of the unit normals to the edges.}
	\label{fig:fig2}
	\end{center}
\end{figure}

We can give a more rigorous geometrical proof of this result, considering the triangle in figure \ref{fig:fig2}, assuming causality and that $E^a \beta_a \leq 0$. The setting is similar to that of the previous geometric argument, however, note that now $(t,x)$ is not an arbitrary inertial frame, but it has been chosen is such a way that $\beta^a \propto \delta\indices{^a _t}$. Furthermore, the arbitrary initial perturbation has now been imposed on the side $C$ of the triangle and not on the x-axis. Again we can apply the Gauss theorem, to obtain
\begin{equation}
E[A]+E[B]+E[C] \leq 0.
\end{equation}
Since there is no perturbation on $A$, we know that $E[A]=0$. Furthermore, given that the normal to $B$ is 
\begin{equation}
n^a[B] = - \dfrac{\beta^a}{\sqrt{-\beta^b \beta_b}} \, ,
\end{equation}
we can use the condition $E^a \beta_a \leq 0$ to show that $E[B] \geq 0$. Hence, we have
\begin{equation}
-E[C] \geq E[B] \geq 0 \, .
\end{equation} 
Noting that $E[C]$ is computed taking the normal to $C$ future-directed, as in figure \ref{fig:fig2}, we conclude that $-E[C]$ quantifies the information contained in $C$. Its positiveness, for any possible choice of initial perturbation on $C$ and for any possible triangle (having the properties described in figure \ref{fig:fig2}), leads to (i) and hence to stability.

\textit{Example 1: perfect fluids -} We conclude the letter with a couple of examples. Consider the information current of a perfect fluid \eqref{Eaperf}, assuming that $\delta \sigma=0$ to first order. Then, the condition of stability in the fluid rest frame reduces to (note that $u^a=T\beta^a$) 
\begin{equation}
-TE^a u_a = (\rho+p) \dfrac{\delta u^b \delta u_b}{2} + \dfrac{ (\delta p)^2}{2(\rho+p)c_s^{2}} \geq 0. 
\end{equation}
This produces the conditions $\rho +p >0$ (positive inertial mass \cite{MTW_book}) and $c_s^2 \geq 0$ (stability of the fluid against compression), which exist also in the Newtonian theory. The causality requirement $E^a E_a \leq 0$ reads
\begin{equation}
(\rho +p)\delta u^b \delta u_b + \dfrac{ (\delta p)^2}{(\rho +p)c_s^{2}} \pm 2 \delta p \sqrt{\delta u^b \delta u_b} \geq 0,
\end{equation}
which produces the well-known condition $c_s^2 \leq 1$ (subluminal speed of sound). The reader might be surprised that $c_s^2 \leq 1$ is also a stability condition. After all, a sound-wave that propagates faster than light is still governed by a wave equation, hence its amplitude should remain bounded over time. However, again we need to remember that a system is \textit{thermodynamically} stable if it is stable also to virtual processes.  One can verify that a virtual process in which the amplitude of a sound-wave grows with time increases the entropy of the fluid in those reference frames in which the sound-wave moves backwards in time, generating instability \cite{GavassinoSuper2021}. Indeed, it is well-known that if a causal microscopic Lagrangian produces an effective macroscopic fluid theory with $c_s^2 >1$, then the equilibrium state is unstable and the perfect fluid description is not applicable, because some high frequency modes must grow \cite{Ruderman1968superlum,Susskind1969,BludmanIstab1970}.

\textit{Example 2: Cattaneo equation -} As a second example we consider a rigid infinite solid bar (1+1 dimensions in flat spacetime), with uniform density, and we model the heat propagation within extended irreversible thermodynamics \cite{Jou_Extended,GavassinoUEIT2021}. We take the fields $(\varphi_i)=(T,q)$, representing temperature and heat flux, as degrees of freedom and impose, in the rest-frame of the solid, the conservation law
\begin{equation}\label{consurvo}
n c_p \partial_t T + \partial_x q =0. 
\end{equation}
The $(t,x)$ components of the entropy current are postulated to be
\begin{equation}\label{sasasa}
s^a = \bigg( \, s - \dfrac{1}{2} \chi q^2 \, , \, \dfrac{q}{T} \, \bigg) \, , 
\end{equation}
where $s=s(T)$ is the equilibrium entropy density. Combining the conservation law \eqref{consurvo} and the constitutive relation \eqref{sasasa}, one can show (just apply the technique of \cite{GavassinoGibbs2021}) that the information current is
\begin{equation}\label{EAAAAA}
E^a = \bigg( \, \dfrac{nc_p (\delta T)^2}{2T^2} + \frac{1}{2}  \chi (\delta q)^2 \, , \, \dfrac{\delta q \, \delta T}{T^2} \bigg).
\end{equation}
The requirement $E^0 > 0$ for $\delta \varphi_i \neq 0$ immediately produces the stability conditions
\begin{equation}
c_p > 0  \spc \chi > 0,
\end{equation}
the first ensuring stability to heat diffusion \cite{PrigoginebookModernThermodynamics2014}, the second to fluctuations of $q$. The requirement that $E^a$ should not be space-like ($E^a E_a \leq 0$) produces 
\begin{equation}
\dfrac{1}{\chi n c_p T^2} \leq 1.
\end{equation}
This is, indeed, the causality condition of the model (but it is also an important stability condition, see \cite{GavassinoLyapunov_2020}, Appendices 3-4). In fact, if we postulate an information annihilation rate $\nabla_a E^a = - (\delta q)^2/(\kappa T^2) \leq 0$ ($\kappa>0$ is the heat conductivity coefficient), the resulting linearised field equation is the Cattaneo equation
\begin{equation}\label{catteno}
\chi n c_p T^2 \, \partial_t^2   T - \partial_x^2  T + \dfrac{n c_p}{\kappa} \partial_t  T =0,
\end{equation}
whose characteristic maximum signal propagation speed is $(\chi n c_p T^2)^{-1/2}$ \cite{Kost2000}. Again, we see that the causality condition is merely thermodynamic (it involves only thermodynamic coefficients) and is unaffected by the value of the kinetic coefficient $\kappa$. In fact, while causality is a geometric constraint on the direction of the information current, $\kappa$ only quantifies the rate at which information is destroyed. In the limit in which $\kappa \rightarrow +\infty$, heat does not propagate infinitely fast. Instead, information becomes a conserved quantity, and \eqref{catteno} becomes a non-dissipative causal wave equation.

\textit{Conclusions -} On the practical side, our work shows that the entropy-based stability criterion developed in \cite{GavassinoGibbs2021} is enough also to ensure linear causality, simplifying the job of testing the reliability of a theory. On the theoretical side, it reveals the central importance of the information current $E^a$ in relativistic hydrodynamics, shedding new light on the role of information theory in a relativistic context. The reason why it took so long to achieve this understanding is that the focus has been up to now on trying to connect causality with \textit{hydrodynamic} stability, while the real connection is with \textit{thermodynamic} stability, which is a much more complete reliability criterion. \\

This work was supported by the Polish National Science Centre grants SONATA BIS 2015/18/E/ST9/00577 and OPUS 2019/33/B/ST9/00942. Partial support comes from PHAROS, COST Action CA16214. LG thanks M. Shokri and G. Torrieri for stimulating discussions. We also thank M. M. Disconzi and the anonymous referees for providing useful comments, which helped us improve the clarity of the paper.

\bibliography{Biblio}

\onecolumngrid

\begin{center}
\Large \bf Supplementary Material
\end{center}

\hspace{2cm}

\hspace{2cm}

\textit{Part 1}: We show that the ``energy currents'' $E^a$ computed from the Gibbs stability criterion \cite{GavassinoGibbs2021} must coincide with the current $E^a$ introduced in equation (3) of the main text. Furthermore, we prove rigorously that $E^a$ coincides with the information current. Finally, we show that $E^a$ can be used to extend the standard theory of thermodynamic fluctuations to account for the fluctuations of the flow velocity in a fully relativistic setting.\\

\textit{Part 2}: We generalize to 3+1 dimensions the proofs for the stability-causality arguments reported in the main text, accounting also for the curvature of space-time.

\section*{Part 1: Uniqueness, thermodynamic origin and statistical meaning of the information current}

\subsection{Assumptions and notation}

We work in the physical setting described in \cite{GavassinoGibbs2021}, adopting also the same notation, according to which $\varphi_i$ are the (macroscopic) fields in equilibrium and $\varphi_i +\delta \varphi_i$ are the fields in a perturbed state. 
For a generic observable $\mathcal{A}$, its \textit{finite} perturbation is defined as
\begin{equation}
\delta \mathcal{A}=\mathcal{A}[\varphi_i + \delta \varphi_i]-\mathcal{A}[\varphi_i].
\end{equation}
The background spacetime is fixed (hence $\delta g_{ab}=0$) and has one and only one independent symmetry generator $K^a$, which is assumed time-like future-directed. In equilibrium it is possible to define the inverse temperature four-vector field $\beta^a=u^a/T$ ($u^a u_a=-1$) and the chemical potential scalar field $\mu$, such that \cite{Becattini2016Beta}
\begin{equation}\label{alfabeto}
\dfrac{\mu}{T}= \alpha  \spc  \beta^a = \beta K^a  \spc \alpha,\beta = \text{const}  \spc \beta>0.
\end{equation}
The complete set of possibly independent conserved (i.e. divergence-free) currents of the system is 
\begin{equation}\label{BasisCons}
\{ \, f(T) \beta^a  \, , \, s^a \, , \, N^a  \, , \, T^{ab}K_b  \, , \, \delta N^a  \, , \, \delta T^{ab} K_b \, \} \, ,
\end{equation} 
where $s^a$, $N^a$ and $T^{ab}$ are entropy current, particle current and symmetric stress-energy tensor. The scalar $f(T)$ is an arbitrary function of the temperature $T=(-\beta^b \beta_b)^{-1/2}$. The conservation of $f(T) \beta^a$, for any $f$, follows from the fact that $\beta^a$ is a Killing vector field. Finally, note that in \cite{GavassinoGibbs2021} all quantities refer to the \textit{total} isolated fluid: there is no separation between fluid and bath. To recover the setting of this letter, we only need to divide the total fluid into two parts (``F'' and ``H'') with infinitely different size.

\subsection{Uniqueness}

All the quadratic ``energy currents'' $E^a$ computed in \cite{GavassinoGibbs2021} have the two following properties:
\begin{itemize}
\item Their total flux across any \textit{Cauchy} 3D-surface is $-\delta S$, provided that $\delta Q^I=0$ for all $I$.
\item They respect conditions (i,ii,iii). 
\end{itemize} 
Let us prove that there can be only one current $E^a=E^a[\varphi_i, \, \delta \varphi_i]$ with these properties. 
To show it, we will assume that there are two such vector fields, $E^a$ and $\tilde{E}^a$, and we will verify that they must be identical.
First, we note that, if the flux of $E^a$ and $\tilde{E}^a$ is $-\delta S$ for \textit{any} Cauchy surface, then (using the Gauss theorem)
\begin{equation}\label{Eaaa}
\nabla_a E^a = \nabla_a \tilde{E}^a =-\nabla_a \, \delta s^a.
\end{equation}
It follows that the difference $z^a :=E^a -\tilde{E}^a$ is a conserved current ($\nabla_a z^a=0$). But since \eqref{BasisCons} is a basis for all the independent conserved currents that we can can build out of $\varphi_i$ and $\delta \varphi_i$, it follows that $z^a$ must be a linear combination of them (with constant coefficients $h_j$):
\begin{equation}
z^a = h_0 f(T) \beta^a +  h_1 s^a +h_2 N^a +h_3 T^{ab} K_b + h_4 \delta N^a + h_5 \delta T^{ab} K_b.
\end{equation}
Condition (ii) implies that wherever the perturbation is absent we must have $E^a=\tilde{E}^a=z^a=0$, hence $h_0 =h_1 = h_2 = h_3 =0$ and we can write
\begin{equation}
z^a =  h_4 \delta N^a + h_5 \delta T^{ab} K_b.
\end{equation}
Finally, we note from condition (i) that, under the transformation $\delta \varphi_i \rightarrow -\delta \varphi_i$, the sign of $E^a$ and $\tilde{E}^a$ cannot change. Hence $z^a$ (like $E^a$ and $\tilde{E}^a$) is a second-order quantity in the perturbations $\delta \varphi_i$. However, both $\delta N^a$ and $\delta T^{ab}$ contain non-zero first-order contributions, so that the only way for $z^a$ to be of second order is that $h_4=h_5=0$. This implies that $z^a=E^a-\tilde{E}^a=0$, completing our proof. 

\subsection{Thermodynamic interpretation}

Since the current $E^a$ (defined in \cite{GavassinoGibbs2021}) is unique, there must be a general thermodynamic formula for it. Here we compute it. From \eqref{Eaaa}, it follows that $w^a:= E^a+\delta s^a$ is a conserved current ($\nabla_a w^a=0$), which again implies
\begin{equation}
w^a =m_0 f(T) \beta^a + m_1 s^a +m_2 N^a +m_3 T^{ab} K_b + m_4 \delta N^a + m_5 \delta T^{ab} K_b  \spc  m_j=\text{const}.
\end{equation} 
Since both $E^a$ and $\delta s^a$ are zero wherever the perturbation is absent, we must impose $m_0 =m_1=m_2=m_3=0$ and we can write
\begin{equation}\label{mapoi}
E^a =-\delta s^a +m_4 \delta N^a + m_5 \delta T^{ab} K_b. 
\end{equation}
This relation is indeed consistent with the condition $\delta s^a = (\text{zfc})-E^a$ reported in \cite{GavassinoGibbs2021}. We are left with the problem of determining the value of the constant coefficients $m_4$ and $m_5$. In order to compute them, we consider a small region of space $\mathcal{R}$ (i.e. a small space-like 3D-surface element) which is locally orthogonal to $K^a$. The particles, energy and entropy contained in $\mathcal{R}$ are
\begin{equation}
\{ N \, , \, U \, , \,S \,  \}= \int_{\mathcal{R}} \{ \, N^a \, , \, - T^{ab} K_b \, ,  \,  s^a \,    \} \, n_a d\Sigma  \spc n^a \text{ past-directed},
\end{equation} 
so that equation \eqref{mapoi}, truncated to the first order in the perturbation (namely neglecting $E^a$), implies
\begin{equation}\label{m4m5}
\delta S = m_4 \delta N-m_5 \delta U. 
\end{equation}
If we work in local inertial coordinates aligned with $K^a$ (and $\mathcal{R}$ is sufficiently small) then,
\begin{equation}
U = - \int_{\mathcal{R}}  T^{ab} K_b n_a d\Sigma = K \int_{\mathcal{R}}  T^{00}  \, d^3 x   \spc  K= \sqrt{-K^b K_b}.
\end{equation}
This implies that $U/K$ is the internal energy (dividing by the red-shift factor $K$ we effectively remove the gravitational potential energy) as measured by a local inertial observer moving with four-velocity $-n^a$, so that from standard thermodynamics we know that (to the first order)
\begin{equation}\label{m7m8}
 \delta S = - \dfrac{\mu}{T} \delta N + \dfrac{\delta U}{KT} .
\end{equation}
Comparing \eqref{m4m5} with \eqref{m7m8}, recalling equation \eqref{alfabeto}, we finally obtain 
\begin{equation}
m_4 = -\dfrac{\mu}{T} = -\alpha  \spc m_5 = -\dfrac{1}{KT} = -\beta.
\end{equation} 
Inserting them into \eqref{mapoi} we have our formula for the information current:
\begin{equation}\label{trovato}
E^a = -\delta s^a -\alpha \delta N^a -\beta_b \delta T^{ab} .
\end{equation}
This formula is not unexpected. In fact, since $E^a$ must be a pure second-order quantity, the first-order truncation of \eqref{trovato} produces Israel's covariant Gibbs relation \cite{Israel_Stewart_1979}:
\begin{equation}
\delta s^a = -\alpha \delta N^a -\beta_b \delta T^{ab} \, .
\end{equation}
However, $N^a$ and $-K_b T^{ab}$ are the conserved currents of the system, whose charges are $N$ and $U$, whose thermodynamic conjugates are $\alpha$ and $-\beta$. Therefore, equation \eqref{trovato} can be rewritten as\footnote{Note that, according to the notation of \cite{GavassinoGibbs2021}, the unperturbed quantities are all evaluated at equilibrium. However, at equilibrium one has $\alpha_I^\text{F}=\alpha_I^\star$ [entropy's maximum: $0=dS_{\text{tot}}=dS_\text{F}+dS_\text{H} =-\alpha^\text{F}_I dQ^I_\text{F}-\alpha^\star_I dQ^I_\text{H}=-(\alpha_I^\text{F}-\alpha_I^\star)dQ^I_\text{F}$]. Therefore, we could write $\alpha=\alpha_{\text{eq}}=\alpha^\star$ and $\beta=\beta_{\text{eq}}=\beta^\star$.} $E^a=-\delta s^a-\alpha_I^\star \delta J^{Ia}$, proving the mathematical equivalence of the Gibbs criterion \cite{GavassinoGibbs2021} with the stability criterion of this letter.

We finally note that, if we multiply \eqref{trovato} by $T$, we are able to define a new current
\begin{equation}
\delta \Omega^a = TE^a = -u_b \delta T^{ab} -T \delta s^a - \mu \delta N^a \, ,
\end{equation}
whose flux across $\mathcal{R}$ is
\begin{equation}
\delta \Omega = \dfrac{\delta U}{K} - T \delta S -\mu \delta N.
\end{equation}
This is nothing but the perturbation to the grand potential of the region $\mathcal{R}$ at fixed $T$ and $\mu$. So, the condition (i), which implies $\delta \Omega \geq 0$, in the end reduces to the statement that the grand potential of small volume elements is minimised in equilibrium \cite{Callen_book}, consistently with the fact that the volume element $\mathcal{R}$ is a subsystem which can exchange energy and particles with the rest of the fluid at temperature $T$ and chemical potential $\mu$. Hence, the fluid elements at equilibrium are not in the maximum entropy state (only the \textit{total} system is in the maximum entropy state), but in the \textit{minimum grand-potential state}.
%But this is nothing but the perturbation to the grand potential of the region $\mathcal{R}$ at fixed $T$ and $\mu$. So, the condition (i), which implies $\delta \Omega \geq 0$, in the end reduces to the statement that the grand potential of small volume elements is minimised in equilibrium \cite{Callen_book}. Also this result is not surprising, as we may regard the volume element $\mathcal{R}$ as a small system which can exchange both energy and particles with a large environment (the rest of the fluid) at temperature $T$ and chemical potential $\mu$. 

\subsection{Current of information}\label{currunz}

Finally, we want to prove that $E^a$ is the current of information carried by the perturbation $\delta \varphi_i$. 

First of all we need to state precisely how we quantify the information. We define our \textit{ignorance} about the state of an \textit{isolated} system as the natural logarithm of the number of microstates in which the system can be, compatibly with our knowledge. We assume that the energy and the number of particles of the system are known to be in the intervals $[U, \, U+\Delta U]$ and $[N, \, N+\Delta N]$, with $\Delta U \ll U$ and $\Delta N \ll N$, so that the maximum possible amount of ignorance is the microcanonical entropy. If we make a measurement of a property of the system, our \textit{ignorance} is reduced by an amount that we call \textit{information}.

Following this line of thoughts, we can define the amount of information $I[\mathcal{R},\, \varphi_i, \, \delta \varphi_i]$ carried by a perturbation $\delta \varphi_i$, contained in a region of space $\mathcal{R}$, as the information that we would gain about the system (about the system as a whole, not just about the region $\mathcal{R}$) measuring all the macroscopic fields $\varphi_i + \delta \varphi_i$ on $\mathcal{R}$, assuming to have no previous knowledge (apart from that of $U$ and $N$ and hence of the equilibrium fields $\varphi_i$, which are microcanonical averages, so they do not constitute additional knowledge). Thus, it follows from the definition that
\begin{equation}\label{Iuuz}
I[\mathcal{R},\, \varphi_i, \, \delta \varphi_i] = -\ln \bigg( \dfrac{\Gamma[U,\Delta U,N,\Delta N, \varphi_i +\delta \varphi_i \text{ on } \mathcal{R}]}{\Gamma[U,\Delta U,N,\Delta N]}  \bigg) \spc \text{with} \spc \Gamma=(\text{number of microstates}) .
\end{equation} 
Now we only need to rewrite the right-hand side as a hydrodynamic integral. We call $\mathcal{R}^c$ (namely, the complementary of $\mathcal{R}$) an arbitrary portion of space such that 
\begin{equation}
\mathcal{R} \cup \mathcal{R}^c = \Sigma  \spc \mathcal{R} \cap \mathcal{R}^c =  \emptyset \, ,
\end{equation}
where $\Sigma$ is a smooth space-like Cauchy 3D-surface. Then we can use Boltzmann's formula for the entropy and make the identifications
\begin{equation}\label{Micruz}
\begin{split}
& \ln \Gamma[U,\Delta U,N,\Delta N, \varphi_i +\delta \varphi_i \text{ on } \mathcal{R}] = S[\mathcal{R}, \varphi_i + \delta \varphi_i] + \max_{\delta \varphi_i \text{ on }\mathcal{R}^c} \, S[\mathcal{R}^c, \varphi_i + \delta \varphi_i] \\ 
& \ln\Gamma[U,\Delta U,N,\Delta N] = S[\Sigma, \varphi_i] \, . \\
\end{split}
\end{equation}
The maximum in the first formula appears because $\Gamma[U,\Delta U,N,\Delta N, \varphi_i +\delta \varphi_i \text{ on } \mathcal{R}]$ constrains only the value of $\delta \varphi_i$ on $\mathcal{R}$, while it sums over all the admissible choices of $\delta \varphi_i$ outside $\mathcal{R}$ (in the thermodynamic limit the configuration $\delta \varphi_i$ that maximizes the entropy dominates the sum \cite{huang_book}). In addition, note that, in the computation of the maximum, we are not completely free to choose $\delta \varphi_i$ on $\mathcal{R}^c$ because the perturbation needs to conserve the total energy and particle number (which are known). Thus we must impose the constraints
\begin{equation}\label{constro}
\delta U[\mathcal{R}^c,\, \varphi_i, \, \delta \varphi_i] = -\delta U [\mathcal{R},\, \varphi_i, \, \delta \varphi_i]=: -\delta U_{\mathcal{R}} \spc \delta N [\mathcal{R}^c,\, \varphi_i, \, \delta \varphi_i] = -\delta N [\mathcal{R},\, \varphi_i, \, \delta \varphi_i]=: - \delta N_{\mathcal{R}} \, .
\end{equation}
Combining \eqref{Iuuz}, \eqref{Micruz} and \eqref{constro}, recalling the formula \eqref{trovato}, we obtain
\begin{equation}\label{IEminE}
I[\mathcal{R},\, \varphi_i, \, \delta \varphi_i] = E[\mathcal{R},\, \varphi_i, \, \delta \varphi_i] + \min_{\delta \varphi_i \text{ on }\mathcal{R}^c} \, E[\mathcal{R}^c,\, \varphi_i, \, \delta \varphi_i].
\end{equation} 
Finally, let us study the second term on the right-hand side. To derive a qualitative upper bound on its typical value (clearly, the minimum of $E$ cannot exceed the value assumed by $E$ on a specific state) we can restrict our attention to configurations $\delta \varphi_i$ on $\mathcal{R}^c$ which are approximately homogeneous across the domain occupied by the fluid, namely 
\begin{equation}\label{nocmke}
(\delta \varphi_i)_{ \text{on }\mathcal{R}^c} \sim \text{const} \, .  
\end{equation}
This is possible only if the theory is causal: in acausal theories the initial data imposed on $\mathcal{R}$ might propagate to $\mathcal{R}^c$ \cite{CourantHilbert1953,Susskind1969,Hishcock1983}, producing unphysical constraints on $\delta \varphi_i$. To estimate the order of magnitude of $(\delta \varphi_i)_{ \text{on }\mathcal{R}^c}$, assuming \eqref{nocmke}, we can use equation \eqref{constro}, considering that $\varphi_i$ are intensive variables, to derive the estimates
\begin{equation}
 V \, (\delta \varphi_i)_{ \text{on }\mathcal{R}}  \sim \{ \delta N_{\mathcal{R}} \, , \, \delta U_{\mathcal{R}} \, \}  \sim - V_c \, (\delta \varphi_i)_{ \text{on }\mathcal{R}^c} \, ,
\end{equation}
with
\begin{equation}
V = \binom{\text{ Volume }}{\text{of }\mathcal{R} } \spc \spc V+ V_c = \binom{\text{Total volume }}{\text{ occupied by the fluid }} \, .
\end{equation}
These estimates, in turn, can be used to show that (recall that $E^a$ is quadratic in the perturbation)
\begin{equation}
V \, E[\mathcal{R},\, \varphi_i, \, \delta \varphi_i] \sim  V^2 \, (\delta \varphi_i)^2_{ \text{on }\mathcal{R}}  \sim V^2_c \, (\delta \varphi_i)^2_{ \text{on }\mathcal{R}^c} \sim V_c \, E[\mathcal{R}^c,\, \varphi_i, \, \delta \varphi_i]\, .
\end{equation}
Hence, we have obtained the qualitative bound
\begin{equation}\label{ventisette}
0 \leq \min_{\delta \varphi_i \text{ on }\mathcal{R}^c} \, E[\mathcal{R}^c,\, \varphi_i, \, \delta \varphi_i] \lesssim \dfrac{V}{V_c} \, E[\mathcal{R},\, \varphi_i, \, \delta \varphi_i] \, ,
\end{equation}
where the first inequality is a consequence of stability. In the limit in which the region $\mathcal{R}$ is infinitely small compared to the size of the whole fluid (namely $V \ll V_c$), equation \eqref{IEminE} reduces to
\begin{equation}
I[\mathcal{R},\, \varphi_i, \, \delta \varphi_i] = E[\mathcal{R},\, \varphi_i, \, \delta \varphi_i] \times \bigg[ \, 1+\mathcal{O}\bigg(  \dfrac{V}{V_c} \bigg) \, \bigg] \approx E[\mathcal{R},\, \varphi_i, \, \delta \varphi_i]= \int_{\mathcal{R}} E^a[\varphi_i, \, \delta \varphi_i]  \, n_a d\Sigma \, ,
\end{equation}
proving that $E^a$ can be interpreted as the current of information.

\subsection{Consistency with the theory of thermodynamic fluctuations}

Let us compute the explicit formula of $E^a$ for perfect fluids, to second order in $\delta \varphi_i$. Take an arbitrary smooth curve in the state-space of F parameterized with a free variable $\epsilon$ [i.e., a set of configurations $\{\varphi_i(\epsilon)\}_{\epsilon \in \mathbb{R}}$ of the fluid]. Consider the current $\phi^a(\epsilon):= s^a(\epsilon)+\alpha^\star N^a(\epsilon)+\beta^\star_b \, T^{ab}(\epsilon)$, where we recall that $\alpha^\star$ and $\beta^\star_b$ are external parameters, whose value is determined by the external conditions (namely, by the bath H). Then, if $\dot{A}:=dA/d\epsilon$, and the fluid is a perfect fluid, we can write (the dependence on the parameter $\epsilon$ is understood)
\begin{equation}
\begin{split}
 \phi^a ={ } & \big[ s+\alpha^\star n + \beta_b^\star u^b (\rho+p) \big]u^a+P\beta^{\star a} \\ 
 \dot{\phi}^a ={ } & \big[ \dot{s}+\alpha^\star \dot{n} + \beta_b^\star u^b (\dot{\rho}+\dot{p})  + \beta_b^\star \dot{u}^b (\rho+p) \big]u^a+ \big[ s+\alpha^\star n + \beta_b^\star u^b (\rho+p) \big]\dot{u}^a +\dot{P}\beta^{\star a} \\
 \ddot{\phi}^a ={ } & \big[ \ddot{s}+\alpha^\star \ddot{n} + \beta_b^\star u^b (\ddot{\rho}+\ddot{p})  + \beta_b^\star \ddot{u}^b (\rho+p) +2\beta_b^\star \dot{u}^b (\dot{\rho}+\dot{p}) \big]u^a+ \big[ s+\alpha^\star n + \beta_b^\star u^b (\rho+p) \big]\ddot{u}^a \\ & +2 \big[ \dot{s}+\alpha^\star \dot{n} + \beta_b^\star u^b (\dot{\rho}+\dot{p})  + \beta_b^\star \dot{u}^b (\rho+p) \big]\dot{u}^a+\ddot{P}\beta^{\star a} \, . \\
\end{split}
\end{equation}
If the curve is such that $\epsilon=0$ is the equilibrium state, then we have
\begin{equation}
E^a (\epsilon) = \phi^a(0)-\phi^a(\epsilon) = -\dfrac{1}{2} \, \ddot{\phi}^a(0)  \, \epsilon^2 + \mathcal{O}(\epsilon^3).
\end{equation}
With the aid of the identities (valid of all values of $\epsilon$)
\begin{equation}
\begin{matrix*}[l]
\rho+p=Ts+\mu n \\
\\
u^b u_b=-1 \\
\end{matrix*}
\spc \spc
\begin{matrix*}[l]
\dot{\rho}=T\dot{s}+\mu \dot{n} \\
\\
u^b \dot{u}_b=0 \\
\end{matrix*}
\spc \spc
\begin{matrix*}[l]
 \ddot{\rho}=T\ddot{s}+\mu \ddot{n} + \dot{T}\dot{s}+\dot{\mu}\dot{n} \\
\\
u^b \ddot{u}_b +\dot{u}^b \dot{u}_b=0 \, ,\\
\end{matrix*}
\end{equation}
we obtain\footnote{Note that all the terms with a second derivative ($\ddot{\varphi}_i$) cancel out, so that the final formula for $E^a$ is quadratic in $\delta \varphi_i$. For this reason, we could just invoke the first-order replacement $\delta \varphi_i(\epsilon) :=\varphi_i(\epsilon)-\varphi_i(0)= \dot{\varphi}_i(0) \, \epsilon +\mathcal{O}(\epsilon^2)$.}
\begin{equation}
\begin{split}
& \mu/T = \alpha^\star  \spc u_b/T = \beta^\star_b  \spc [\text{equilibrium conditon: }\dot{\phi}^a(0)=0] \\
& TE^a = \bigg[ \delta T \, \delta s + \delta \mu \, \delta n  + (\rho+p)\delta u^b \delta u_b \bigg]\dfrac{u^a}{2} + \delta p \, \delta u^a +\mathcal{O}(\epsilon^3) . \\
\end{split}
\end{equation}
Using the thermodynamic relations
\begin{equation}
dp=s \, dT+n \,d\mu \spc dT=\dfrac{T}{c_p} d\sigma +\dfrac{Tk_p}{nc_p} dp  \spc d\bigg(\dfrac{1}{n}\bigg) = \dfrac{Tk_p}{nc_p} d\sigma-\dfrac{dp}{n(\rho+p)c_s^2},
\end{equation}
where our notation is summarised in the main text (except for $k_p$, which is the isobaric thermal expansivity), we find that, retaining terms up to second order in $\epsilon$,
\begin{equation}
 \delta T \, \delta s + \delta \mu \, \delta n =n \big[ \delta T \, \delta \sigma - \delta p \, \delta(n^{-1}) \big] = \dfrac{nT(\delta \sigma)^2}{c_p} + \dfrac{(\delta p)^2}{c_s^2(\rho+p)} \, ,
\end{equation}
so that we recover equation (7) of the main text. Assuming that F is homogeneous in its support, $\delta N_\text{F}=0$, and $\Sigma$ is such that $n^a=-u^a$, we can rewrite equation (6) of the main text (the probability distribution for fluctuations) as follows:
\begin{equation}
\mathcal{P}[\delta S_\text{F}, \delta p,\delta u^b]\propto \exp \bigg[ -\dfrac{(\delta S_\text{F})^2}{2C_p} -\dfrac{V(\delta p)^2}{2Tc_s^2 (\rho+p)} - \dfrac{V(\rho+p)}{2T}\delta u^b \delta u_b \bigg].
\end{equation}
This generalizes equation (15) of Section 15.1 of \citet{Pathria2011}, accounting for fluctuations of the flow velocity.

\newpage

\section*{Part 2: Proof of the stability-causality arguments in 3+1 dimensions}

\subsection{The argument for causality}

\begin{figure}[h!]
\begin{center}
\includegraphics[width=0.8\textwidth]{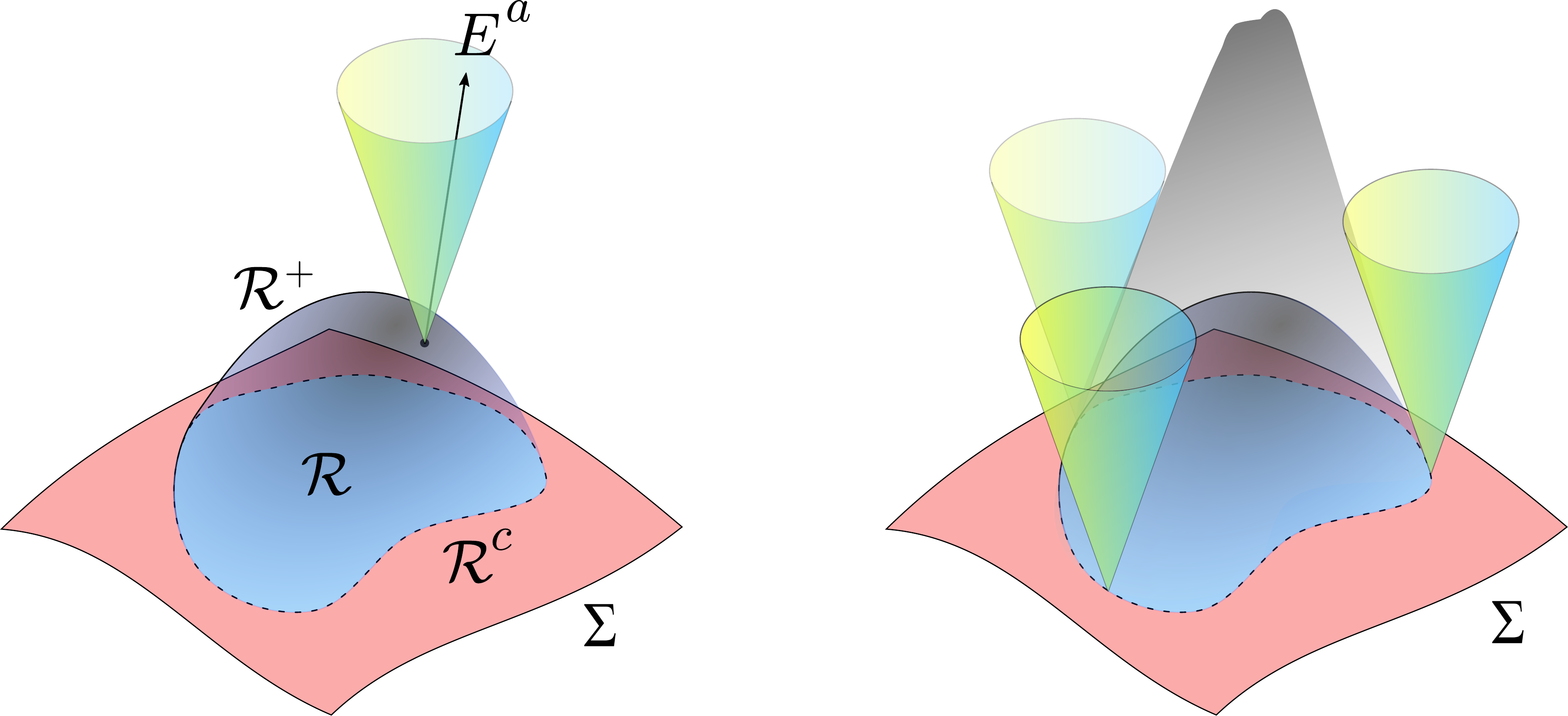}
	\caption{Visualization of the geometric argument. The Cauchy surface $\Sigma$ is divided into an equilibrium region $\mathcal{R}$ (blue), where $\delta \varphi_i=0$, and a perturbed region $\mathcal{R}^c$ (red), which surrounds $\mathcal{R}$, where $\delta \varphi_i \neq 0$. If the dynamics is causal, the perturbation cannot enter the future Cauchy development of $\mathcal{R}$ (the shaded ``pyramid'' in the right panel). In fact, the fluid elements contained in $\mathcal{R}$ ``cannot know'' that the region $\mathcal{R}^c$ is out of equilibrium \textit{before} the perturbation reaches them; hence, they will evolve \textit{as if} the fluid were in global thermodynamic equilibrium. To prove it, we show that the information current $E^a$ vanishes on any space-like ``dome'' $\mathcal{R}^+$.}
	\label{fig:fig3D}
	\end{center}
\end{figure}

We consider a space-like Cauchy 3D-surface $\Sigma$, which is the disjoint union of two space-like surfaces $\mathcal{R}$ and $\mathcal{R}^c$, as in figure \ref{fig:fig3D}. We assume that $\mathcal{R}$ is compact (it represents a finite region of space) and that the perturbation $\delta \varphi_i$ vanishes on $\mathcal{R}$. Therefore, we know from condition (ii) that
\begin{equation}\label{zeroonR}
E^a =0 \quad \quad \text{on }\mathcal{R} \, .
\end{equation}
By contrast, we assume that the fluid is out of equilibrium on $\mathcal{R}^c$, so that $E^a \neq 0$ on $\mathcal{R}^c$. ``Causality'' means that the perturbation on $\mathcal{R}^c$ cannot propagate inside the future Cauchy development $\mathcal{D}^+(\mathcal{R})$ of $\mathcal{R}$, because the latter lies outside the domain of influence of $\mathcal{R}^c$ (see figure \ref{fig:fig3D}, right panel) \cite{Hawking1973}. Hence, what we need to show is that
\begin{equation}\label{goaAAAAAAAAAAll}
E^a =0 \quad \quad \text{on }\mathcal{D}^+(\mathcal{R}) \, .
\end{equation} 
In order to do it, let us consider an arbitrary \textit{space-like} 3D-surface $\mathcal{R}^+ \subset \mathcal{D}^+(\mathcal{R})$, whose two-dimensional spatial boundary coincides with that of $\mathcal{R}$ (visually, one can imagine $\mathcal{R}^+$ as a sort of ``dome'', covering $\mathcal{R}$ entirely\footnote{If we compare the present analysis with figure 1 of the main text (which was restricted to 1+1 dimensions), $\mathcal{R}$ is the analogue of the lower side $C$ of the triangle, while the union of the two upper sides (namely $A \cup B$) is a particular choice of $\mathcal{R}^+$.}). The set $\mathcal{R} \cup \mathcal{R}^+$ is a closed orientable surface, whose interior may be called $I(\mathcal{R} \cup \mathcal{R}^+)$. Applying the Gauss theorem, and recalling condition (iii), we obtain
\begin{equation}\label{inequo}
\int_{\mathcal{R} \cup \mathcal{R}^+} \!\!\!\!\!\!\!\!\! E^a n_a \, d\Sigma = \int_{I(\mathcal{R} \cup \mathcal{R}^+)} \!\!\!\!\!\!\!\!\!\!\!\!\! \nabla_a E^a \, d\mathcal{V} \leq 0 \, .
\end{equation}
On the other hand, equation \eqref{zeroonR} implies that the surface integral over $\mathcal{R} \cup \mathcal{R}^+$ coincides with that over $\mathcal{R}^+$, so that \eqref{inequo} becomes
\begin{equation}\label{leSecunda}
\int_{\mathcal{R}^+ }  E^a n_a \, d\Sigma  \leq 0 \, .
\end{equation}
Now, let us determine the sign of $E^a n_a$ on $\mathcal{R}^+$. For the Gauss theorem to be valid, as given in equation \eqref{inequo}, the sign of the one-form $n_a$ must be chosen in such a way that \cite{DrayDivergence1994,Hawking1973,FengConserved2018}
\begin{equation}\label{gbuz}
E^a n_a >0  \spc \text{if }E^a\text{ is out-ward pointing} \, .
\end{equation}
On the other hand, we know from condition (i) that $E^a$ is time/light-like future-directed. Therefore, it lies inside (or belongs to) the future light-cone, which points out of the space-time region $I(\mathcal{R} \cup \mathcal{R}^+)$ on $\mathcal{R}^+$ (see also figure \ref{fig:fig3D}, left panel, and recall that $\mathcal{R}^+$ is space-like). Therefore,
\begin{equation}\label{elgringo}
E^a n_a \geq 0 \quad \text{on }\mathcal{R}^+ \, .
\end{equation} 
The only way for \eqref{leSecunda} and \eqref{elgringo} to hold simultaneously is that 
\begin{equation}
E^a =0 \quad \quad \text{on }\mathcal{R}^+ \, .
\end{equation}
On the other hand, the interior of the future Cauchy development of $\mathcal{R}$ can be foliated with ``domes'' like $\mathcal{R}^+$, so that \eqref{goaAAAAAAAAAAll} follows.

\subsection{The inverse argument}

To prove the inverse argument, one needs to consider a situation that is essentially ``specular'' to that of figure \ref{fig:fig3D}. Assume that $\mathcal{R}$ is a (not necessarily Cauchy) space-like 3D-surface, such that 
\begin{equation}\label{condizion}
E^a n_a \geq 0  \quad \quad n^a\text{ unit normal to }\mathcal{R}  \quad (n^a \text{ time-like past-directed}) \, ,
\end{equation} 
for arbitrary $\delta \varphi_i$. This just means that the fluid is stable for an observer having four-velocity $u^a =-n^a$. Assuming causality, together with condition (iii), our goal is to prove that $E^a$ is also time-like future-directed on $\mathcal{R}$.

\begin{figure}
\begin{center}
\includegraphics[width=0.5\textwidth]{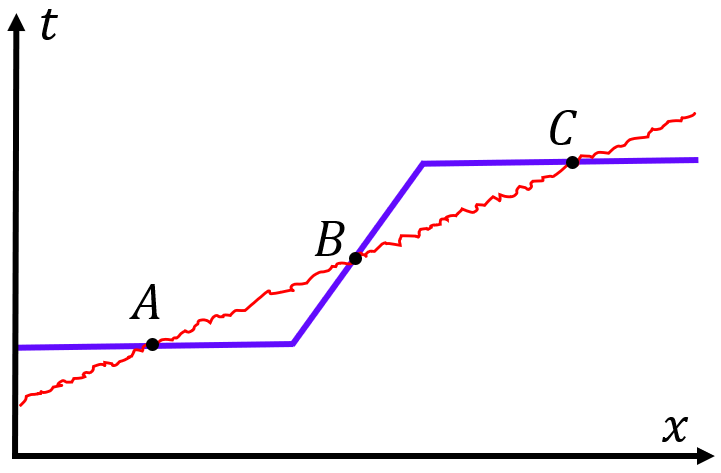}
	\caption{In acausal theories, a travelling signal (red line) can intersect a space-like 3D-surface $\Sigma$ (blue line) more than once. When this happens, information propagates, e.g., from $A$ to $B$, and we are not allowed to set our data on $\Sigma$ freely, but we need to make sure that $\delta \varphi_i(A)$ and $\delta \varphi_i(B)$ are appropriately correlated. For example, if the linearised field equations of the hydrodynamic model read $v^a \nabla_a (\delta \varphi_i)=0$, where $v^a$ is a space-like vector field, tangent to the red line, then we know that $\delta \varphi_i(A)=\delta \varphi_i(B)$. If the state on $\Sigma$ has been constructed in a way that $\delta \varphi_i(A) \neq \delta \varphi_i (B)$, the history of the system is self-contradictory, making such state impossible. Obviously, these issues cannot arise in causal theories, because signals must travel along causal world-lines, which cannot intersect space-like surfaces more than once.}
	\label{fig:infoLINE}
	\end{center}
\end{figure}

We consider a space-like 3D-surface $\mathcal{R}^- \subset \mathcal{D}^-(\mathcal{R})$ (here, $\mathcal{D}^-(\mathcal{R})$ is the \textit{past} Cauchy development of $\mathcal{R}$), whose two-dimensional spatial boundary coincides with that of $\mathcal{R}$ (visually, $\mathcal{R}^-$ looks like a ``cup'', closed from above by $\mathcal{R}$~\footnote{If we compare the present analysis with figure 2 of the main text (which was restricted to 1+1 dimensions), $\mathcal{R}$ is the analogue of the upper side $B$ of the triangle, while the union of the two lower sides (namely $C \cup A$) is a particular choice of $\mathcal{R}^-$.}). Analogously to what we did in the previous subsection, we can apply the Gauss theorem to the closed surface $\mathcal{R} \cup \mathcal{R}^-$, obtaining
\begin{equation}\label{inequo325}
\int_{\mathcal{R} \cup \mathcal{R}^-} \!\!\!\!\!\!\!\!\! E^a n_a \, d\Sigma = \int_{I(\mathcal{R} \cup \mathcal{R}^-)} \!\!\!\!\!\!\!\!\!\!\!\!\! \nabla_a E^a \, d\mathcal{V} \leq 0 \, .
\end{equation} 
Recalling that the sign of the one-form $n_a$ in \eqref{inequo325} is determined by equation \eqref{gbuz}, we can rewrite \eqref{inequo325} as follows:
\begin{equation}\label{erminus}
E[\mathcal{R}^-] \geq E[\mathcal{R}] \geq 0 \, \spc (\text{with unit normal past-directed}) ,
\end{equation}
where the second inequality is a consequence of \eqref{condizion}. Therefore, we have just shown that the functional $E[\mathcal{R}^-]$ is positive definite for any allowed choice of $\delta \varphi_i$. On the other hand, the principle of causality, in its standard definition \cite{CourantHilbert1953,Susskind1969,Hishcock1983}, implies that we are allowed to set the value of $\delta \varphi_i$ on $\mathcal{R}^-$ freely (in acausal theories this is no longer true, because space-like surfaces can intersect characteristic surfaces more than once, see figure \ref{fig:infoLINE}), so that \eqref{erminus} becomes
\begin{equation}\label{inexquo}
E^a n_a \geq 0  \quad \quad n^a\text{ unit normal to }\mathcal{R}^-  \quad (n^a \text{ time-like past-directed}) \, .
\end{equation}
Now, let us pick up an arbitrary point $x \in \mathcal{D}^-(\mathcal{R})^{\mathrm{o}}$. We can always construct $\mathcal{R}^-$ in such a way that $x \in \mathcal{R}^-$, and we can always ``twist'' $\mathcal{R}^-$ so that $n^a$ points in any time-like past direction we like. Since equation \eqref{inexquo} must hold on $x$ for all possible choices of $\mathcal{R}^-$ passing through $x$, we recover (i) on $x$. In conclusion, condition (i) is valid on $\mathcal{D}^-(\mathcal{R})^{\mathrm{o}}$. By continuity, it must remain valid also on $\mathcal{R}$, completing our proof.

\newpage

\subsection{Information is non-local in acausal theories}

The role of causality in our proof for the ``inverse argument'' given above is to covert the \textit{global} condition $E[\mathcal{R}^-] \geq 0$ into the \textit{local} condition $E^a n_a \geq 0$. Let us see in more detail why the principle of causality is crucial for making this passage possible.

Let us consider, for clarity, a perfect fluid, with $c_s >1$. We assume that $\delta \varphi_i$ has the form of a small wave-packet of sound-waves, travelling superluminally across the fluid. Working the the fluid's rest frame, we are interested in computing the total flux of $E^a$ across a space-like Cauchy 3D-surface $\Sigma$, having the shape reported in figure \ref{fig:infoLINE}. This choice of surface is particularly interesting, because according to an observer located on $B$, with four-velocity normal to $\Sigma$ (recall that $\Sigma$ is space-like, hence $n^a(B)$ is time-like), the sound-wave is moving \textit{backwards} in time.

The total flux $E[\Sigma]$ naturally splits into three contributions coming from the three intersections between the wave-packet and $\Sigma$:
\begin{equation}
E[\Sigma]= E_A +E_B +E_C \, .
\end{equation} 
On the other hand, a perfect fluid is non-dissipative ($\nabla_a E^a =0$), hence we can apply the Gauss theorem to the space-time region between $\Sigma$ and a generic time-slice $t=\text{const}$, to show that
\begin{equation}
E[\Sigma] =E[t=\text{const}]= E_A = E_C \, .
\end{equation}
Combining the two equations above, we obtain
\begin{equation}\label{eunz}
E_B =-E_A \, .
\end{equation}
We immediately see the problem: the contribution to $E$ has a  different sign, depending on whether the sound-wave is moving forward or backwards in time, in the reference frame defined by the four-velocity $-n^a$. This can also be verified explicitly: a sound-wave which propagates in the positive $x$ direction (in the fluid's rest frame) is a perturbation $\delta \varphi_i$ such that
\begin{equation}
\delta \sigma = 0 \spc \spc \delta u^1 = \dfrac{\delta p}{c_s (\rho +p)} \, .
\end{equation}
If we plug these condition into equation (5) of the main text we get
\begin{equation}
TE^a = T
\begin{pmatrix}
E^0   \\
E^1 \\
\end{pmatrix}
= \dfrac{(\delta p)^2}{c_s^2(\rho+p)}
\begin{pmatrix}
1   \\
c_s \\
\end{pmatrix} \, .
\end{equation}
Now, if we make a boost of speed $v$ in the $x-$direction, we obtain
\begin{equation}
TE^{a'}  = \dfrac{\gamma(\delta p)^2}{c_s^2(\rho+p)}
\begin{pmatrix}
1-vc_s   \\
c_s -v \\
\end{pmatrix} 
\spc \gamma = \dfrac{1}{\sqrt{1-v^2}} \, .
\end{equation} 
We see that, depending on the value of $v$, the sign of $E^{0'}$ can change. In particular, $E^{0'} < 0$ if and only if $v>c_s^{-1}$, which is what we wanted to verify.

The present discussion has interesting implications in the context of information theory. Consider again our proof that $E^a$ can be interpreted as the information current (subsection \ref{currunz}). We proved equation \eqref{IEminE} on general grounds, but then we needed to invoke the principle of causality to put an upper bound on the second term on the right-hand side, see equation \eqref{ventisette}. Equation \eqref{eunz} clearly shows us that such upper bound is not valid in acausal theories. Therefore, in acausal theories $E^a$ \textit{cannot} be interpreted as the information current. The interpretation of this fact is simple: if the theory is acausal, measuring a property of the system in the region $\mathcal{R}$ gives us some information about the state of the system in $\mathcal{R}^c$ (besides its total energy and particle number). Hence, information is no longer a local quantity and cannot have an associated current.

\end{document}